\documentclass{epl}

\usepackage{epsfig}

\title{Optical properties of crystals with spatial dispersion --
\newline 
 Josephson plasma resonance in layered superconductors }
\shorttitle{Optics with spatial dispersion}
\author{L.N. Bulaevskii\inst{1} \and Ch. Helm\inst{1,2} \and
 A.R. Bishop\inst{1} \and M.P. Maley\inst{1}}
\institute{\inst{1} Los Alamos National Laboratory - Los Alamos, NM 87545 \\
 \inst{2} ETH H{\"o}nggerberg, Institut f{\"u}r Theoretische Physik -
 8093 Z{\"u}rich, Switzerland}
\shortauthor{L.N. Bulaevskii, Ch. Helm {\it et al.}}
\date{\today}


\begin{document}

\maketitle

\begin{abstract}
We derive the transmission coefficient, $T(\omega)$, for grazing incidence of
crystals with spatial dispersion accounting for the excitation of 
multiple modes with different wave vectors ${\bf k}$ for a given frequency 
$\omega$.  The generalization of the Fresnel formulas contains 
the refraction indices of these modes as determined by the dielectric function 
$\epsilon(\omega,{\bf k})$.  Near 
frequencies $\omega_e$, where the group velocity vanishes, 
$T(\omega)$ depends also on an additional parameter determined by the 
crystal microstructure. The transmission $T$ is significantly suppressed, if 
one of the excited modes is decaying into the crystal. We derive these 
features microscopically for the Josephson plasma resonance in layered 
superconductors.
\end{abstract}

Usually propagation of light in crystals is sensitive to average properties 
described by the dielectric function, but not to the specific atomic structure 
of the crystal, because the wave length of light is much larger than the 
interatomic distance.  However, in crystals 
with spatial dispersion the atomic structure may affect optical properties at 
special extremal frequencies, $\omega_e$, where the group velocity, 
${\bf v}_g=\partial \omega({\bf k})/\partial{\bf k}$, of an eigenmode with 
dispersion $\omega({\bf k})$ vanishes. 
Near $\omega_e$ the effective wave length, $\lambda_g=
v_g/\omega$, becomes comparable with the interatomic distance, and then 
optical properties such as the reflectivity may feel the crystal
 microstructure. 

Pekar \cite{pekar} and Ginzburg \cite{agr}, already realized  that several 
eigenmodes with different wave vectors, ${\bf k}$, at given 
$\omega$ may be excited by the incident light inside a crystal with 
spatial dispersion.  Then the Maxwell boundary conditions (continuity 
of the components of the electric, ${\bf E}$, and of the 
magnetic field, ${\bf H}$, parallel to the surface), are not sufficient to 
find the relative amplitudes of these modes and hence the reflection 
coefficient.  To overcome this problem within 
the macroscopic approach Pekar and Ginzburg introduced so called additional 
boundary conditions (ABC), which are supposed to be related somehow 
to the crystal microstructure near the surface.  
In this phenomenological approach the 
choice of these ABC is only heuristic and may be controversial, 
see ref.~\cite{henn} and Comments 
to this paper.  It is only 
a microscopic model which can determine the solutions inside 
the crystal unambiguously and demonstrate explicitly how the reflectivity
 depends on the crystal microstructure. 

In this paper we present for the first time 
general macroscopic expressions for the 
transmission coefficient $T(\omega)$ in uniaxial crystals in the 
vicinity of $\omega_e$ to 
show how the Fresnel formula is modified by the presence of multiple 
solutions. 
Then we find a full microscopic 
description of optical properties for the Josephson plasma resonance (JPR) 
in layered superconductors, which supports the phenomenological picture 
and derives the ABC. 
We show explicitly that the maximum $T(\omega)$ reached at 
$\omega\approx \omega_e$ depends on the interlayer distance of the 
superconductor.   Weak dissipation 
as well as a perfect crystal structure are necessary to observe this effect of 
spatial dispersion, as otherwise the influence of the microstructure is 
smeared out and the conventional Fresnel results are restored.
For details of the calculation see \cite{uslong}.

We also point out that the stopping of light, i.e. ${\bf v}_{g} =0$, 
which in the case of gaseous media  attracted considerable interest for the 
coherent optical information storage recently \cite{stoplight}, 
here naturally appears in a solid at a finite wave vector due to spatial 
dispersion.

The JPR is an interlayer charge oscillation due to tunneling 
Cooper pairs and quasiparticles in layered superconductors.  It is strongly 
underdamped because the quasiparticles responsible for dissipation 
are frozen out at low temperatures. The JPR is an appropriate phenomenon to 
study 
the effect of spatial dispersion on optical properties both theoretically and 
experimentally. Firstly, the quite simple Lawrence-Doniach model 
\cite{art,lnb}  is sufficient to provide a complete 
microscopic description.  Secondly, weak dissipation may be achieved 
at low temperatures if the JPR frequency is well 
below the superconducting gap. 
Thirdly, 
there is experimental evidence that the spatial dispersion of the Josephson 
plasmon is significant.  Recently 
in the layered superconductor SmLa$_{1-x}$Sr$_{x}$CuO$_{4-\delta}$ two 
peaks at $\approx 7$ and $\approx 12$ cm$^{-1}$ were observed 
in optical properties \cite{marel}. 
These peaks can be naturally understood as the JPR of alternating 
intrinsic junctions with SmO or La$_{1-x}$O$_x$ in the barriers
between the CuO$_2$-layers, and the peak intensities indicate a  strong 
inter-junction coupling \cite{art,lnb,mareltheory,helm,sendai}, 
i.e. a significant spatial dispersion of the plasma modes. 
Tachiki et al. \cite{art}  mentioned the excitations of multiple modes and
Marel et al. \cite{mareltheory} suggested an effective dielectric function 
for alternating junctions, 
but in neither case the effects on the reflectivity  presented below 
were discussed. 

To formulate the general problem posed by spatial dispersion on a 
macroscopic level  we consider a layered crystal with 
the dielectric functions $\epsilon_a(\omega)$ along the layers and 
$\epsilon_c(\omega,k_z)$  along the $z$-axis.  In 
$\epsilon_c(\omega,k_z)$
we account for a collective 
mode (here JPR) with the dispersion $\omega_c(k_z)$, i.e. 
$\epsilon_c[\omega_c(k_z),k_z]=0$.  
The incident polarized light of frequency $\omega$ has a magnetic field 
${\bf B}=(0,B_y,0)$ and a wave vector 
${\bf k}_0=(\omega \sin\theta /c,0, \omega \cos\theta/ c)$, where $\theta$ 
is the angle relative to the $z$-axis normal to the surface and to the 
layers.  The bulk Maxwell equations 
 \begin{equation}
c k_xB_y=-\omega\epsilon_c(\omega,k_z)E_z, \; \;  
ck_zB_y=\omega \epsilon_a(\omega)E_x, \; \; 
k_xE_z-k_z E_x=-(\omega/c)  B_y, \label{max} 
\end{equation}
determine the wave vector ${\bf k}=(\omega \sin\theta /c,0,k_z)$ 
 of the eigenmodes at the given frequency $\omega$. Here $k_z(\omega)$
is a solution of the equation
 \begin{equation}
 k_z^2=\epsilon_a(\omega)(\omega^2/c^2)[1-\sin^2\theta/
 \epsilon_c(\omega,k_z)]. 
 \label{dr0}
 \end{equation}
When $\omega_c$ and $\epsilon_c$ are $k_z$-independent, eq.~(\ref{dr0}) has
one solution 
for $k_z^2$ and the Maxwell boundary conditions lead to the conventional 
Fresnel formula  $T=1-|1-\kappa|^2/|1+\kappa|^2$
with $\kappa=n/[\epsilon_a(\omega)
\cos\theta]$, where $n=ck_z(\omega)/\omega$ is the refraction index. Then 
$T(\omega)$ is peaked  near  $\omega_c$. 

In a crystal with spatial dispersion in $\epsilon_c(\omega,k_z)$, 
Eq.~(\ref{dr0}) has multiple solutions  for $k_z^2(\omega)$, 
in the simplest case four (real or complex) ones, $\pm n_1(\omega), 
\pm n_2(\omega)$ \cite{pekar,agr}, which can interfere with each other 
\cite{uslong}.    
For a semi-infinite crystal at $z>0$ only two of them are 
physical.  At low dissipation the energy flow (Poynting vector ${\bf S}$) of
proper modes should be directed into the crystal in order to preserve 
causality.  Thus their group velocity, $v_{gz}(\omega)=\partial 
\omega(k_z)/\partial k_z \sim S_z$,  should be positive \cite{agr}.  
In the case of normal (anomalous) dispersion this requires positive 
(negative) ${\rm Re} (k_z)$.  When dissipation is taken into account, 
this rule is equivalent 
to the condition that the eigenmodes decay inside the crystal, 
${\rm Im}(k_z)>0$.  Without dissipation one gets $v_{gz}=0$ at the frequency 
$\omega_e$, separating the normal and anomalous parts of the dispersion 
curve $\omega(k_z)$.  
At $\omega=\omega_e$, the two physical solutions for $k_z$ have the 
same $|k_z|$, but opposite signs, i.e. $n_1(\omega_e)+n_2(\omega_e)=0$. 

To determine the relative amplitudes of the two modes excited by incident 
light, ABC for the macroscopic  polarization ${\bf P}$ were used in 
\cite{pekar,agr}.  
In the bulk $P_z$ is related linearly to $E_z$ by the 
susceptibility $\chi_c(k_z)=[\epsilon_c(\omega,k_z)-\epsilon_{c0}]/4\pi$, 
 $\epsilon_{c0}$ being the high frequency dielectric constant. 
Generally this relation may differ near the surface. 
This difference is very small for the JPR, as it is confined between adjacent 
superconducting layers both in the bulk and on the surface, and the surface 
intrinsic junction is almost the same as that in the bulk. 
Therefore we restrict ourselves to the dielectric functions $\epsilon_c 
(z,z^\prime) = \Theta (z) \Theta(z^\prime) \epsilon_c (z-z^\prime) $ with a 
sharp cutoff at the surface $z=0$. 

The ABC for our system, as proposed by Ginzburg, 
 \begin{equation}
 P_z(z)+\ell(\partial P_z/\partial z)=0, \ \ \ z\rightarrow 0, 
 \label{ABC}
 \end{equation}
determine the amplitudes $\gamma_{1,2}$  of the fields 
 \begin{eqnarray}
&& E_z(z)=\gamma_1\exp(ik_{z1}z)+\gamma_2\exp(ik_{z2}z), \\
&& P_z(z)=\gamma_1\chi_c(k_{z1})\exp(ik_{z1}z)+
 \gamma_2\chi_c(k_{z2})\exp(ik_{z2}z). 
 \end{eqnarray}
%  \begin{equation}
%E_z(z)=\sum_{p=1,2} \gamma_p\exp(ik_{zp}z), \; \; 
% P_z(z)=\sum_{p=1,2} \gamma_p \chi_c(k_{zp})\exp(ik_{zp}z)
% E_z(z)=\gamma_1\exp(ik_{z1}z)+\gamma_2\exp(ik_{z2}z), \; \; 
% P_z(z)=\gamma_1\chi_c(k_{z1})\exp(ik_{z1}z)+
% \gamma_2\chi_c(k_{z2})\exp(ik_{z2}z). 
% \end{equation}
Here $\ell$ is a phenomenological parameter related to the 
crystal microstructure and is to be determined in a microscopic model.  
Usually, 
$\chi_c(k_z)-\chi_c(0)\propto k_z^2$ at $k_z\rightarrow 0$. In this limit 
the relation for the amplitudes is 
\begin{equation}
\gamma_1(1+i\xi n_{1})+\gamma_2(1+i\xi n_{2})=0, \ \ \ \xi=\omega\ell/c.
\label{ga}
\end{equation}
Using eqs.~(\ref{max})-(\ref{dr0}) and (\ref{ga}), we derive 
\begin{equation}
\kappa=\frac{1}{\cos\theta}~\frac{1+
\epsilon_a^{-1}n_{1}n_{2}}{n_{1}+n_{2}-
i\xi\epsilon_a(1+\epsilon_a^{-1} n_{1}n_{2})}.
\label{gen}
\end{equation} 
In the limit $|n_2|\gg |n_1| \gg \sqrt{| \epsilon_a |} $ the 
refraction index with smallest $|n_p|$ 
 determines $\kappa$ and this condition defines the 
frequencies where the one-mode Fresnel description is correct.  
In the two-mode frequency interval $T(\omega)$ is peaked when $|n_1|=|n_2|$, 
and two situations are possible. 

When both $n_{1,2}$ are real without dissipation we have a frequency 
$\omega_e$, where $n_1+n_2= 0$.  At $\omega=\omega_e$ only the 
term $i \xi n_1n_2$ with $\xi\ll 1$ remains in the denominator   
leading to $T(\omega_e)=0$.  
As $\omega$ increases above $\omega_e$, the transmissivity  
increases and reaches its maximum, 
\begin{equation}
T_{\rm max}=T(\omega_{e, \rm max})=2/[(1+\epsilon^2_a \xi^2 
\cos^2\theta)^{1/2}+1],
\label{F}
\end{equation}
at $\omega=\omega_{e,{\rm max}}$ 
when $n_1+n_2=(1+\epsilon_a^{-1}n_1n_2)(\cos^{-2}\theta+
\epsilon_a^2 \xi^2)^{1/2}$ above $\omega_e$. 
We see that $T_{\rm max}$ depends on $\ell$ and is generally 
smaller than the Fresnel result $T_{\rm max}=1$ without dissipation.  
Note, that the opposite signs  of the refraction indices 
$n_{1,2}$ near $\omega_e$ due to causality are essential for the dependence 
of $T_{\rm max}$ on the length $\ell$. 
The vanishing of $n_1 + n_2$ at $\omega_e$ in eq.~(\ref{gen}) 
and its consequence on $T_{\rm max}$ have 
not been realized previously \cite{pekar,agr,henn,forst}.  

We anticipate an even stronger suppression of $T_{{\rm max}}$ due to spatial 
dispersion at frequencies $\omega_i$ near $\omega_c$, where $n_1$ 
is real, while $n_2=in_1$ without dissipation.  
This occurs in superconductors when the dispersion of the collective mode is 
anomalous.  $T(\omega)$ is peaked 
at $\omega_i$, but $T(\omega_i)\ll T(\omega_{e,{\rm max}})$ (eq.(7)), as 
 now $n_1+n_2\neq 0$ in the denominator of eq.~(\ref{gen}). The microstructure 
$\sim \xi$ is irrelevant at 
$\omega_i$. This observation is confirmed below for the JPR. 

The frequencies $\omega_e$ and $\omega_i$, the frequency interval 
of the two-mode regime near $\omega_e$ and $\omega_i$, 
as well as $n_{1,2}$ are determined  by the  dielectric function, but 
$T_{{\rm max}}$ near $\omega_e$ and $\omega_i$ also depend on the ABC. 
These features and the ABC will now be derived microscopically for 
the JPR. 
 
This requires the solution of the Maxwell equations 
and the equations for the phase of the superconducting order parameter, 
exactly accounting for the  atomic (layered) structure of the crystal 
along the $z$-axis,  supercurrents inside the
2D layers at $z=ms$ and for interlayer Josephson and quasiparticle currents 
between neighboring layers. We obtain 
\begin{eqnarray}
&&c\partial_z B_y =
i\epsilon_{a0}\omega\left[E_x-\frac{\omega_{a0}^2}
{\omega^2}\sum_{m=0}E_xs \delta (z-ms)\right], \label{first} \\
&& \partial_z E_x -ik_xE_z=i\frac{\omega}{c} B_y, 
E_{z,m,m+1}=
\int_{ms}^{(m+1)s} \; \; \; E_z\frac{dz}{s}, \\
&&ck_xB_y=-\omega\epsilon_{c0}\left[
E_z-\sum_{m=0}A_mf_{m,m+1}(z)\right], \label{e} \\ 
&& \frac{{\tilde \omega}_l^2 es}{\omega_{0l}^2} A_m= V_{m,m+1} 
=esE_{z,m,m+1}+\mu_{m+1}-\mu_{m}. \label{last}
\end{eqnarray}
Here  $\mu_m$ is the chemical potential in the layer $m$, 
$V_{m,m+1}$ is the difference of the electrochemical potentials, 
$\omega_{0l}^2=8\pi^2csJ_l/\epsilon_{c0}$ are the bare JPR plasma frequencies
determined by the Josephson critical current densities $J_l$ in junctions of
type $l=1,2$, 
$\omega_{a0}=c/\lambda_{ab}\sqrt{\epsilon_{a0}}$ is the in-plane plasma 
frequency, $\epsilon_{a0}$ is the high frequency in-plane dielectric 
constant and $\lambda_{ab}$ is the in-plane London penetration length.
The function $f$ is defined as
$f_{m,m+1}(z)=1$ at $ms<z<(m+1)s$ and zero elsewhere.
To obtain eq.~(\ref{e}) for small amplitude
oscillations we expressed the supercurrent density
$J_{m,m+1}^{(s)}=J_l\sin\varphi_{m,m+1}\approx J_l\varphi_{m,m+1}$ 
via the phase difference $\varphi_{m,m+1}=2iV_{m,m+1}/\hbar\omega$.
Further, 
$\tilde{\omega}_l^2=\omega^2(1-i4\pi\sigma_{cl} \omega / \omega_{0l}^2
\epsilon_{c0}^{\phantom 2})^{-1}$ takes into account 
the dissipation due to quasiparticle tunneling currents, 
$J_{m,m+1}^{(qp)}=\sigma_{cl}V_{m,m+1}/es$, determined by the 
conductivities $\sigma_{cl}$. 
We express the difference $\mu_m-\mu_{m+1}$  
 via the difference of the 2D charge densities, ${\rho}_m$, as 
$\mu_m-\mu_{m+1}=(4\pi  s\alpha/\epsilon_{c0})({\rho}_m-{\rho}_{m+1})$, 
where the 
parameter $\alpha=(\epsilon_{c0}/4\pi e s)(\partial \mu/\partial {\rho})$ 
characterizes the inter-junction coupling, i.e. the JPR charge dispersion 
\cite{art}.  This parameter $\alpha$ is estimated as 0.4 
for SmLa$_{1-x}$Sr$_{x}$CuO$_{4-\delta}$ \cite{helm}.  Finally, 
the Poisson equation  expresses ${\rho}_m$ via $E_z$.  

The solution of eqs.(\ref{first}) - (\ref{last}) 
between the layers $m$ and $m+1$ is 
 \begin{equation}
 E_z(z)=c_m\exp(igz)+d_m\exp(-igz)+aA_m, 
 \end{equation}
 and similar for $B_y$ and $E_x$, where 
 $g=(\omega^2\epsilon_{a0}/c^2a)^{1/2}$ and 
$a^{-1}= 1-\sin^2\theta/\epsilon_{c0}$.  
The continuity of $B_y$ and $E_x$ across the layers leads to 
a set of finite-difference equations for $c_m$, $d_m$ and $A_m$.  

\begin{figure}
\begin{center}
%\onefigure{schematicmixing2layerletter.ps}
\epsfig{file=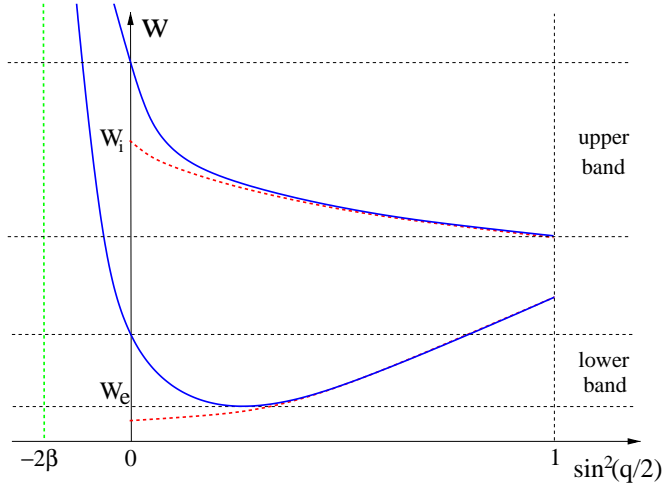,width=0.45\textwidth,angle=-90,clip=}
\end{center}
\caption{
Schematic picture of the dispersion $w= \omega^2 / \omega^2_{0}$ for two 
alternating junctions ($\sigma_c=0$, solid line). The lower band is 
analogous to the case of identical junctions and its
 dispersion is normal  for  $\theta=0$ (dashed).
  Its mixing with a decaying electromagnetic wave 
(as shown  by the dashed, 
vertical line at $\sin^2(q/2) = - 2 \beta $) 
results in two propagating modes with real $q$ near the lower band edge $w_e$, 
where the group velocity vanishes. The anomalous dispersion 
in the upper band gives rise to one propagating and one  
decaying mode and a special point $w_i$, where $n_2 = i n_1$.
\label{schematicmixing} 
}
\end{figure}  

First we consider a crystal with identical junctions, 
$\omega_{0l}=\omega_{0}= c/\lambda_c\sqrt{\epsilon_{c0}}$ and 
$\sigma_{cl}=\sigma_c$.  
Omitting the terms of order $(gs)^2\approx (s/\lambda_c)^2\sim 10^{-10}$ 
we obtain in the bulk
\begin{equation}
A_m(\tilde{w}-a-2\alpha)+\alpha(A_{m+1}+A_{m-1}) = 
(1-2\alpha\beta ) ( c_m+d_m),\label{e3}
\end{equation}
for $A_m$ at $m\geq 1$ and similar equations for $c_m$, $d_m$.  Here 
$\tilde{ w}=\tilde{\omega}^2/\omega_{0}^2$ and   
$\beta=s^2/2\lambda_{ab}^2a$ ($\sim 10^{-4}$ in cuprates).  
Eliminating $c_m$, $d_m$ we find linear equations for $A_m$ alone, which give 
the dispersion of the eigenmodes.  The difference between the equations for 
$A_0$ and $A_1$ is the microscopic boundary condition 
which allows us to find $A_m$ using bulk equations. 
The dispersion of the eigenmodes ($0\leq q\leq 2\pi$), 
\begin{equation}
 w(q)=\frac{\omega^2(q)}{\omega_{0}^2}=
 1+2\alpha(1-\cos q) +\frac{(a-1)\beta}{\beta+1-\cos q} , 
\label{d1}
\end{equation}
is equivalent to eq.~(\ref{dr0}) with 
\begin{equation}
 \epsilon_c(\omega,q)=
\epsilon_{c0}\left[1-\omega_c^2(q) / \omega^2  + 4 \pi i \sigma_c / \omega
\epsilon_{c0} \right], \label{maxwellaverages}  \; \; \;  
 \epsilon_{a}(\omega)=\epsilon_{a0} 
\left[ 1-\omega_{a0}^2 / \omega^2 \right],
\end{equation}
where $
\omega_c^2(q)=\omega_{0}^2[1+\alpha s^2 k_z^2]$ and
  $k_z^2=2(1-\cos q)/s^2$.  The 
dispersion of the plasma  mode due to the charging effect,
 $\omega_c(q)$, is normal at low $q$ \cite{art}. 
The last term in eq.~(\ref{d1}) describes the
anomalous (at small $q$) dispersion due to the inductive coupling of the 
in-plane currents \cite{lnb}.  When alone, it describes the decay of an 
electromagnetic 
wave inside the superconductor on the length $\lambda_{ab}$.  
As shown in fig.~1 for the lower band, their combination leads to 
the extremal frequency, $w_e=1+u$,   
with $u= [8(a-1)\beta\alpha]^{1/2}$, if $\alpha>
(a-1)\beta/8$ and if the dissipation is weak, 
$4\pi\sigma_c/\omega_{0}\epsilon_{c0}  \ll u$. 
For any angle $\theta \neq 0$ 
the longitudinal plasma oscillations mix 
effectively with the electromagnetic wave at small 
$q\approx \sqrt{u/2\alpha}$. 
From eq.~(\ref{d1}) we find 
$n_{1,2}^2=(c/\omega s)^2[\tilde{w}-1\pm \sqrt{(\tilde{w}-1)^2-u^2}]/2\alpha$ 
and 
$|n_1n_2|=\lambda_c^2\epsilon_{c0}u/2\alpha s^2
\approx \lambda_c^2/(s\lambda_{ab})\gg 1$.  
Hence, the two-mode regime, when $|n_1|\sim|n_2|$,  occurs in the 
frequency interval of width
$\approx u\omega_{0}$ above $\omega_e$.  Otherwise, at 
$|\tilde{w}-1|\gg u$, the standard Fresnel formulas are correct.

Note that by modifying the JPR frequency 
$\omega_0$ and hence $\omega_e$, e.g. by an 
external magnetic field, the stopping of light at $\omega_e$ 
can be modified fastly, which might serve as the building block of a 
future magnetooptical device.

Next we find solutions for $A_m$, $c_m$, $d_m$.  We obtain for $m\geq 1$
\begin{equation}
A_m=\gamma_1 A(q_1)\exp(iq_1m)+\gamma_2 A(q_2) \exp(iq_2m) . 
\label{Aq}
\end{equation}
The microscopic boundary condition found as described above is 
$A_{-1}=0$, where  $A_{-1}$ is the extension of the bulk
solution, eq.~(\ref{Aq}), to $m=-1$.  
This  corresponds simply to the fact that for the surface 
 junction, $m=0$, one neighboring junction, $m=-1$, is missing.
We note that 
$A_m$ plays the role of the polarization, $P_z$, 
because it describes the response of the system  
to the electrochemical potential, see eq.~(\ref{last}). With 
$A_{-1}=P_z(z=-s)$  we derive microscopically eq.~(\ref{ABC}) with 
 $\ell=-s$. Then in eq.~(\ref{F}) we get $\xi\epsilon_a=
s\lambda_c\sqrt{\epsilon_{c0}}/\lambda_{ab}^2$ which may be of order unity in 
cuprates like Tl-2212 with $\lambda_c/\lambda_{ab}\sim 100$ 
and the JPR frequency $\sim 20$ cm$^{-1}$.  

Hence, for grazing incidence spatial dispersion ($\alpha\neq 0$) 
leads to (i) a shift of the peak position  
in $T(w)$ by $u  \sim \sqrt{\beta}$, and (ii) a
decrease of $T_{{\rm max}}$ depending on the interlayer spacing $\sim \xi$,
see fig.~2. 

In a crystal with  alternating Josephson junctions, 
 $\delta=\omega_{01}^2/\omega_{02}^2< 1$, we introduce $A_1(n)=A_{2n}$ and 
$A_2(n)=A_{2n+1}$.  They describe the polarization inside two different 
junctions in the unit cell $n$.  For a plasmon  
decoupled from the electromagnetic field (at $\theta=0$), 
the charge coupling of junctions leads to the dispersion of the two 
 bands  and a gap between them, see fig.~1.    
In the lower (upper) band the dispersion is normal (anomalous).  
Mixing with the decaying 
 electromagnetic mode in the lower band leads 
to the existence of an extremal frequency $\omega_e$ for 
$\alpha^2>(a-1)\beta(1-\delta)/4$ and 
$4 \pi \sigma_{0l}/\omega_{01} \epsilon_{c0}  \ll u$. 
Hence, the situation in the lower band is similar to 
that in the case of identical junctions. 
In contrast to this, the  mixing of the upper plasma band with 
anomalous dispersion and the decaying electromagnetic mode leads to an  
eigenmode with anomalous dispersion everywhere.  
Here $n_1$ is real, while $n_2$ is imaginary, and a critical frequency 
$\omega_i$ exists, see fig.~1.   

\begin{figure}
\begin{center}
%\onefigure{Tw.ps} 
\epsfig{file=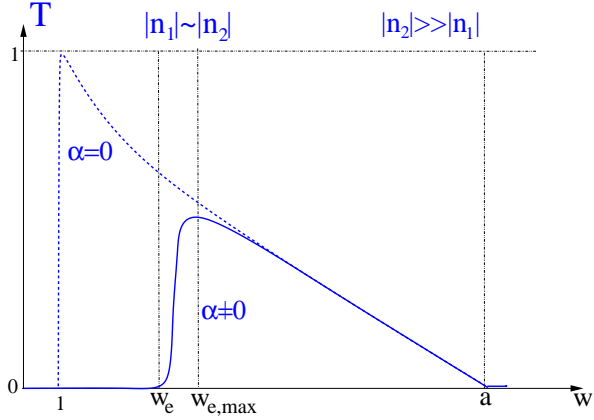,width=0.45\textwidth,angle=-90,clip=}
\end{center}
\caption{Schematic transmission coefficient $T(w)$ 
with (solid) and without (dashed) spatial dispersion ($\sigma_c = 0$). 
As $\alpha$ increases, the peak in $T(w)$ is shifted to higher frequency 
and its amplitude decreases depending on the interlayer spacing 
$\sim \xi$ ($w = \omega^2/ \omega_0^2$, $\alpha$ charge 
coupling). 
\label{RTschematic} 
}
\end{figure}  

When a junction of type 1 is at the boundary, the missing neighboring 
 junction  is of  type 2 and the microscopic boundary condition is 
derived as $A_2(n=-1)=0$. 
The solution of the equations for $A_1(n)$ and $A_2(n)$, i.e. the 
generalization of eqs.~(\ref{e3}) and (\ref{Aq}) for alternating 
junctions, are the eigenvectors
$[A_1(q_p), A_2(q_p)]$, $(p=1,2)$, which describe the microstructure of the
local polarization within the unit cell for each mode $p$. Here we obtain 
\begin{equation}
\kappa=\frac{1}{\cos\theta}~
\frac{[1+\epsilon_a^{-1}n_1n_2][{\tilde A}_2(q_1)n_2-{\tilde A}_2(q_2)n_1]}
{{\tilde A}_2(q_1)n_2^2-{\tilde A}_2(q_2)n_1^2} , 
\end{equation}
where ${\tilde A}_2(q)=A_2(q) \exp (-iq) \approx 
A_2(0)(1-iq/2)$ at small $q=2sk_z$. 
This leads  in the lower band to eq. (\ref{gen}) for $\kappa$
and a similar behavior of $T(w)$ as in the case of identical junctions.  
In the upper band the atomic scale  $\xi$ is irrelevant.
There the maximum value of $T(w)$ is reached at $w_{{\rm max}}=w_i$ when 
 $n_1= in_2$ for $\sigma_{0l}=0$, 
 \begin{equation}
T_{{\rm max}}\approx \frac{2 \lambda_{ab}^{3/2}\epsilon_{a0}}
 {\lambda_c(s\epsilon_{c0})^{1/2}\cos\theta }
 \left[\frac{(a-1)L}{8\alpha^2 a}\right]^{1/4}.
 \end{equation}
Here $L=w_i(1+\delta)-2-8\alpha$, $w_i=(1+\delta)(1+2\alpha)(1+
\sqrt{1-p})/2\delta$ is 
the upper edge of the 
band at $a=1$, $q=0$ and 
$p=4\delta(1+4\alpha)/(1+\delta)^2(1+2\alpha)^2$.  Thus $T_{\rm max}$ depends 
on $\alpha$ explicitly and  is smaller 
by  the factor $(s/\lambda_{ab})^{1/2}$ than 
$T_{{\rm max}}$ in the lower  band or for the case with identical junctions,
respectively.  This 
difference vanishes in the Fresnel limit, when 
$4  \pi \sigma_{0l}\gg \omega_{0l}u$. 

In conclusion, multiple eigenmodes, propagating or decaying, 
are excited by incident light in crystals 
with spatial dispersion.  
At low dissipation this leads to a drop of the maximum transmission  
near resonance frequencies in comparison with the single-mode Fresnel result. 
In the case of two propagating modes 
near frequencies $\omega_e$, where the group velocity vanishes, 
the maximum transmission coefficient, $T_{{\rm max}}$, depends on the 
crystal microstructure, which is in remarkable contrast to the general belief
that atomic scales cannot affect optical properties due to the large
wavelength of light. 
The drop of $T_{{\rm max}}$ is larger when one 
excited mode is propagating, while the other is decaying.  
This behavior was demonstrated  explicitly for the JPR, 
but it is general for any optically active 
excitations, e.g. optical phonons with anomalous dispersion in insulators.  
However, the condition of weak 
dissipation and a perfect crystal structure are crucial to observe 
deviations from the Fresnel regime. 

\acknowledgments
The authors thank Dirk van der Marel for useful discussions.  
This work was supported by the US DOE 
and the Swiss National Science Foundation
through the National Center of Competence
in Research "Materials with Novel Electronic Properties-MaNEP".

\end{document}